\documentclass[a4paper,11pt]{article}
\usepackage{pos}

\title{Cosmic ray energy spectra derived from KASCADE-Grande data using post-LHC hadronic interaction models}
\ShortTitle{KASCADE-Grande energy spectra}

\author{
D.~Kang$^{a,*}$,   
J.C.~Arteaga-Vel\'azquez$^{b}$,
M.~Bertaina$^{c}$,
A.~Chiavassa$^{c}$,
A.L.~Colmenero-C\'esar$^{b}$,
K.~Daumiller$^{a}$,
V.~de Souza$^{d}$,
R.~Engel$^{a,e}$,
A.~Gherghel-Lascu$^{f}$,
C.~Grupen$^{g}$,
A.~Haungs$^{a}$,                                              
J.R.~H\"orandel$^{h}$,                                                         
T.~Huege$^{a}$,                                                                
K.-H.~Kampert$^{i}$,                                                           
K.~Link$^{a}$,                                                                 
H.J.~Mathes$^{a}$,                                                             
S.~Ostapchenko$^{j}$,                                                          
T.~Pierog$^{a}$,                                                               
D.~Rivera-Rangel$^{b}$,                                                        
M.~Roth$^{a}$,                                                                 
H.~Schieler$^{a}$,                                                             
F.G.~Schr\"oder$^{a}$,                                                         
O.~Sima$^{k}$,                                                                 
A.~Weindl$^{a}$,                                                               
J.~Wochele$^{a}$,                                                              
J.~Zabierowski$^{l}$}
\affiliation[a]{Karlsruhe Institute of Technology, Institute for Astroparticle Physics, Germany}
\affiliation[b]{Universidad Michoacana, Instituto de F\'{i}sica y Matem\'{a}ticas, Morelia, Mexico}
\affiliation[c]{Dipartimento di Fisica, Universit\`{a} degli Studi di Torino, Italy}
\affiliation[d]{Universidade S\~{a}o Paulo, Instituto de F\'{i}sica de S\~{a}o Carlos, Brasil}
\affiliation[e]{Karlsruhe Institute of Technology, Institute of Experimental Particle Physics, Germany} 
\affiliation[f]{Horia Hulubei National Institute of Physics and Nuclear Engineering, Bucharest, Romania}
\affiliation[g]{Department of Physics, Siegen University, Germany}
\affiliation[h]{Dept. of Astrophysics, Radboud University Nijmegen, The Netherlands}
\affiliation[i]{Fachbereich Physik, Universit\"at Wuppertal, Germany}
\affiliation[j]{Hamburg University, II Institute for Theoretical Physics, 22761 Hamburg}
\affiliation[k]{Department of Physics, University of Bucharest, Bucharest, Romania}
\affiliation[l]{National Centre for Nuclear Research, Department of Astrophysics, Lodz, Poland}
\affiliation[*]{\rm Speaker}

\emailAdd{donghwa.kang@kit.edu}

\abstract{KASCADE-Grande was dedicated to measuring the energy spectrum and mass composition of cosmic rays in the energy range of 10 PeV to 1 EeV. We observed a knee-like structure in the heavy mass component at around 100 PeV and an ankle-like structure in the light component. In this contribution, we present updated energy spectra based on shower size measurements, using the post-LHC hadronic models QGSJet-II-04, EPOS-LHC, and SIBYLL 2.3d, including accounting for shower-to-shower fluctuations. In addition, the newly released EPOS-LHC-R model is tested for the first time with KASCADE-Grande. We will compare and discuss the results obtained using the different hadronic interaction models.}

\ConferenceLogo{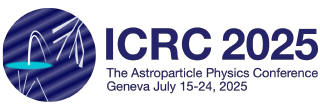}

\FullConference{39th International Cosmic Ray Conference (ICRC2025)\\
 15–24 July 2025\\
Geneva, Switzerland\\}

\begin{document}
\maketitle

\section{Introduction}
To understand the transition between galactic and extragalactic cosmic rays, experiments investigate the energy spectrum and mass composition of cosmic rays in the range from PeV to EeV. 
KASCADE~\cite{KASCADE} and its extension KASCADE-Grande~\cite{KASCADE-Grande} were dedicated to studying extensive air showers in this energy range, particularly their primary energy, chemical composition, and arrival direction.
Both experiments were located at the Karlsruhe Institute of Technology in Germany (49.1$^{\circ}$N, 8.4$^{\circ}$E, 110 m a.s.l.) and measured extensive air showers produced by cosmic rays in the PeV to EeV energy range. 
It is necessary to note that KASCADE's multi-detector system enabled separate measurements of the number of electrons and muons in each shower, providing for studying mass composition.

Over more than 20 years of data collection, the experiments revealed valuable results.
KASCADE observed a "knee" at around 3~PeV in the all-particle energy spectrum, mainly due to a drop in the number of light particles, such as protons and helium nuclei~\cite{KA_APP}. 
KASCADE-Grande extended the energy range and observed a "concave" feature in the spectrum above 10$^{16}$ eV as well as an "iron-like" break around 10$^{17}$ eV. 
This second break likely results from a steepening in the spectrum of heavy elements such as iron, which fits with the expectation that heavier particles are less deflected in magnetic fields~\cite{KG_APP, KG_PRL}. 
Additionally, KASCADE-Grande found an "ankle-like" feature around 100~PeV in the spectrum of light cosmic rays~\cite{KG_PRD}. This feature might indicate a sign of an extragalactic component, which already dominates below $10^{18}$ eV.

Although the data collection completed in 2013, analysis continues. Among others, the data are now used to test and improve models of cosmic-ray interaction in the atmosphere. 
In this paper, we present updated energy spectra of light and heavy cosmic ray components, 
along with corrections for shower fluctuations based on an unfolding method.
We also discuss recent studies on a new version of the hadronic interaction model EPOS.

\section{All-particle and individual energy spectra}

This work analyzes approximately 17 million events recorded by KASCADE-Grande over a total measuring time of 1865.62 days.
After applying quality cuts including zenith angles of up to 40$^{\circ}$, only events with full trigger and reconstruction efficiency were considered. Showers with more than 10$^{6}$ charged particles correspond to a primary energy of approximately 10$^{16}$~eV.

Air shower simulations were performed using the CORSIKA program~\cite{CORSIKA} with different hadronic interaction models. FLUKA was used for low-energy (E < 200~GeV) interactions, while QGSJET-II-04~\cite{QGS04}, EPOS-LHC~\cite{EPOS}, SIBYLL 2.3d~\cite{SIB23d}, and EPOS-LHC-R~\cite{EPOSR} were used for high-energy interactions. The simulations included five types of primary nuclei: proton, helium, carbon, silicon, and iron. Simulated showers cover a primary energy range from 10$^{14}$ to 10$^{18}$~eV and zenith angles from 0$^{\circ}$ to 42$^{\circ}$.
They were initially generated with a spectral index of -2 but subsequently reweighted to -3 for analysis.

The all-particle energy spectrum, along with individual spectra for heavy and light primary groups, is reconstructed based on the shower size ($N_{ch}$), which represents the number of charged particles.
Measured data is divided into two mass groups, heavy and light, on an event-by-event basis.
This separation is based on the correlation between the attenuation-corrected number of charged particles and the number of muons, referred to as the parameter 
$y_{CIC} = {\rm log}_{10}(N_{\mu})^{CIC}_{CF} / {\rm log}_{10}(N_{ch})^{CIC}$.
Hadronic interaction models predict different values for the shower size and muon numbers at the ground level.
Consequently, the $y_{CIC}$ values are model-dependent.  
In general, air showers induced by heavy primaries, such as silicon and iron nuclei, contain fewer electrons at ground compared to those induced by light primaries, such as protons, helium and carbon nuclei. As a result, heavy primaries are referred to as the "electron-poor" group, while light primaries are defined as the "electron-rich" group.

Energy 
calibration functions
are obtained by performing a linear fit to the correlation between the true primary energy and the shower size, both plotted on logarithmic scales. The slope of the fit is approximately 1.
Further details on the energy calibration coefficients can be found in Ref.~\cite{KANG}.
Using these energy calibration relations, the spectra for heavy and light primary mass group are reconstructed accordingly.

Since shower fluctuations in the energy determination are larger than the bin size of the reconstructed energy spectrum, an unfolding procedure is applied to correct for these effects.
A brief summary of the method is provided below. More details can be found in Ref. \cite{KG_APP}. 
Using Monte Carlo simulations, a response matrix $R_{ij}$ is constructed over the full energy interval of log$_{10}(E/\rm GeV)$ from 6.0 to 9.0, where fluctuations can affect the measured energy spectrum. 
This matrix represents the conditional probability $P(E_{j}|E^{true}_{i})$, i.e., the probability that 
an event with true energy in bin log$_{10}(E^{true}_{i})$ is reconstructed with energy log$_{10}(E_{j})$.
Based on this response matrix, a system of equations,
$n^{exp}_{j} = \sum_{i=1}^{N} P(E_{j}|E^{true}_{i}) n^{true}_{i}$,
is established to relate the measured events $n^{exp}_{j}$ to the true energy distribution, $n^{true}_{i}$.
This system is solved iteratively using the Bayesian unfolding algorithm~\cite{Bayes}.
The impact of this unfolding procedure on the resulting flux is estimated to be less than 10\% for all energy bins and therefore does not significantly alter the overall shape of the spectra. 

\begin{figure}[t]
    \centering
    \includegraphics[width=0.329\linewidth]{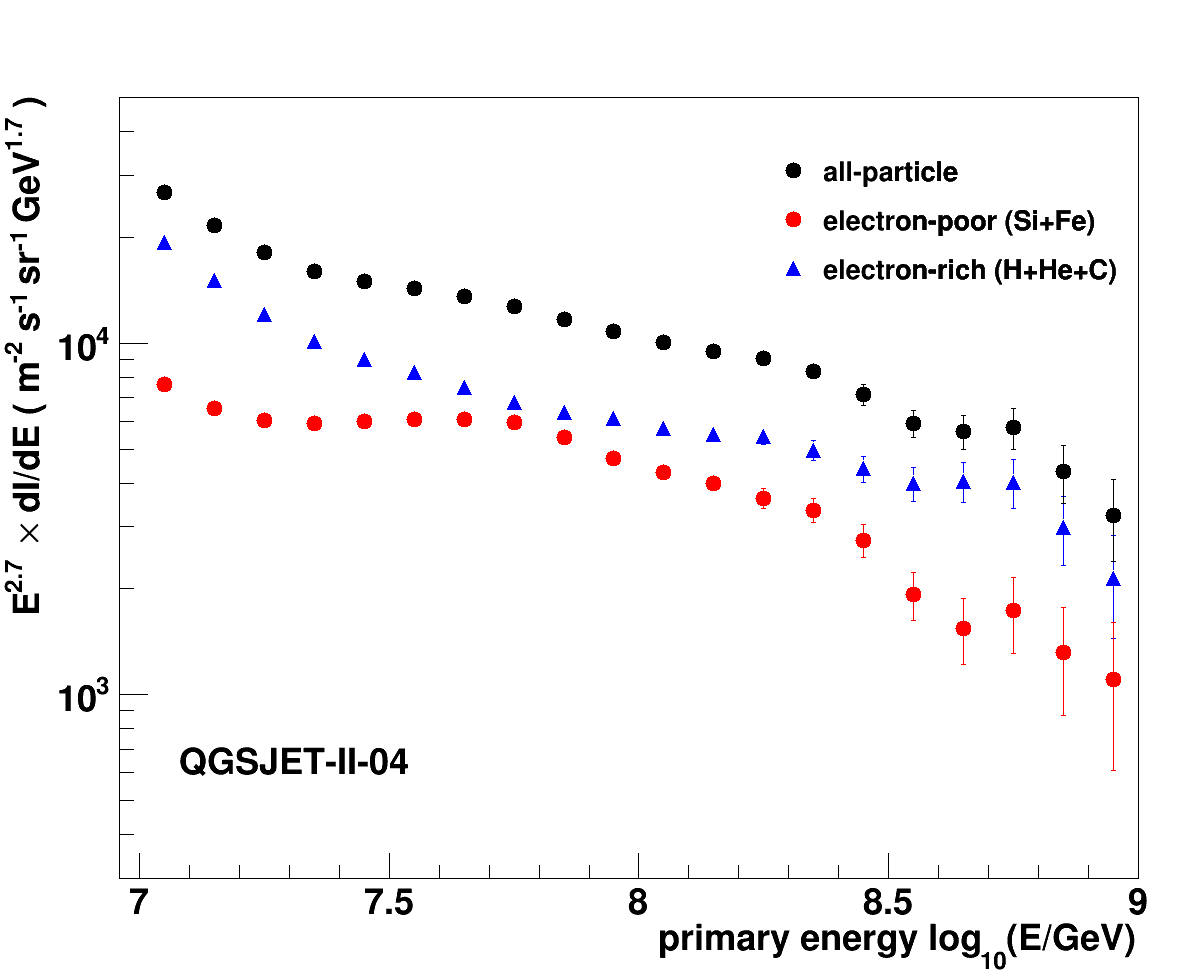}
        \includegraphics[width=0.329\linewidth]{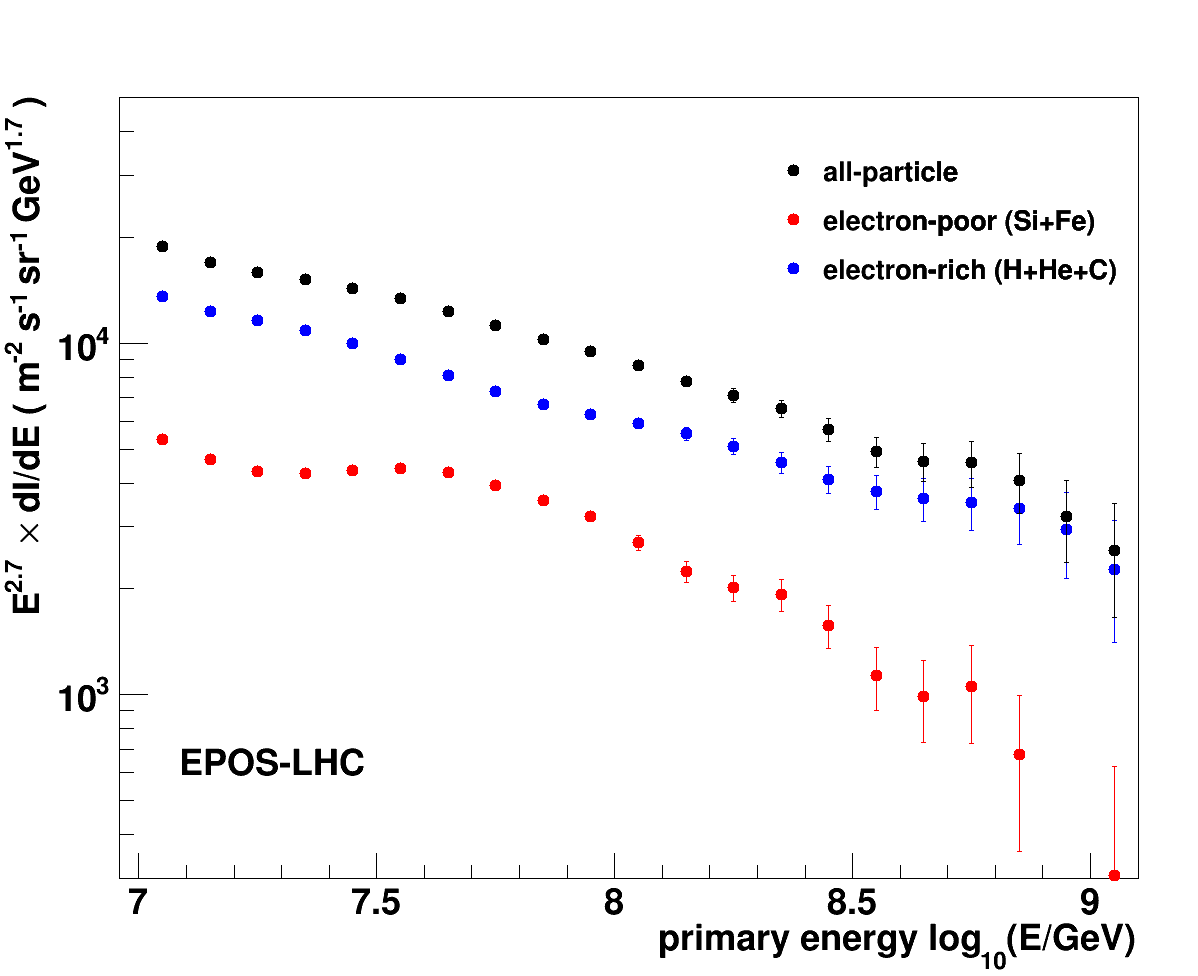}
            \includegraphics[width=0.329\linewidth]{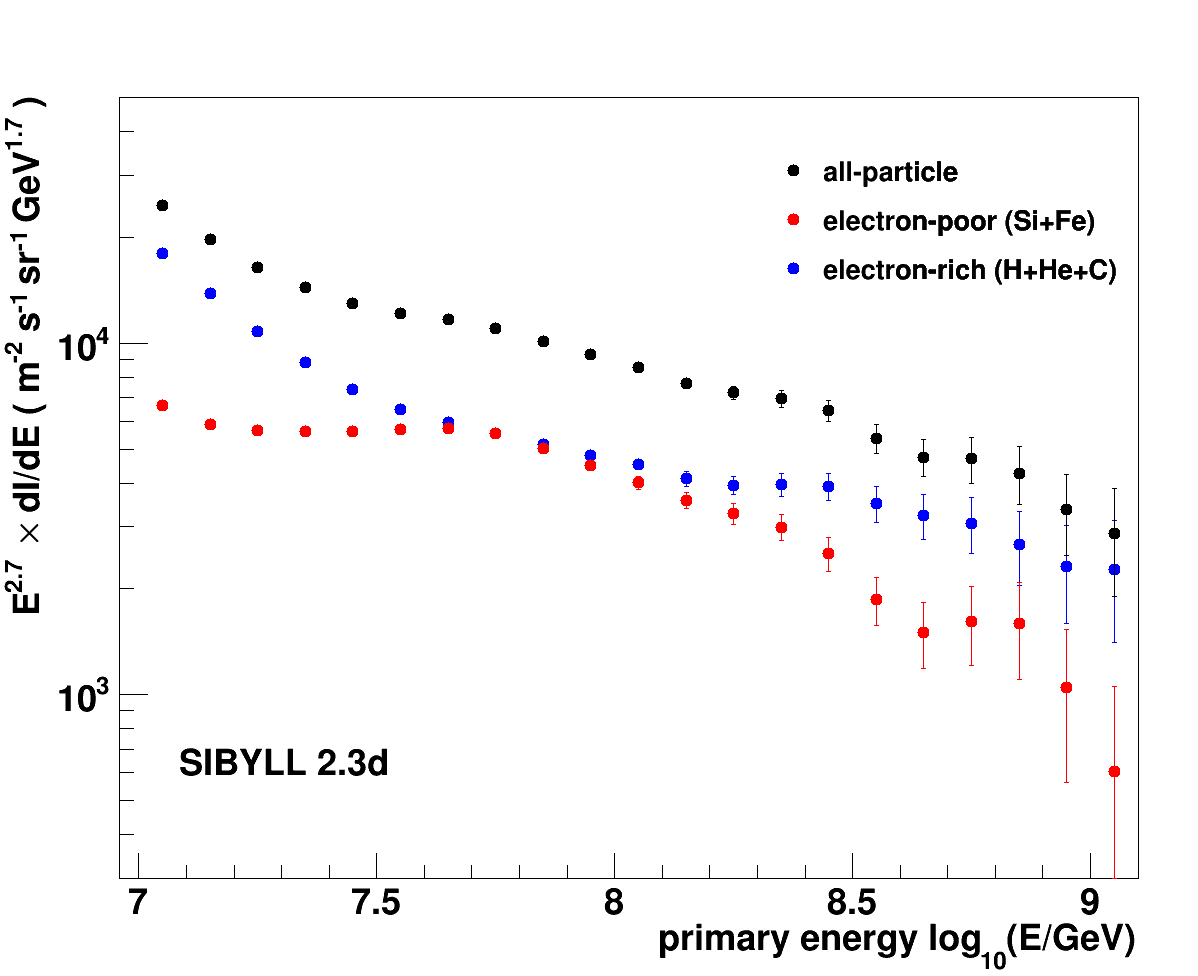}
    \caption{Reconstructed energy spectra for heavy (Si+Fe) and light (p+He+CNO) primary mass groups based on QGSJET-II-04 (left), EPOS-LHC (middle), and SIBYLL 2.3d (right), with shower-to-shower fluctuation corrections applied. The all-particle is the sum of the light and heavy mass compositions. Only statistical uncertainties are shown, for a discussion on systematics see Ref.~\cite{KANG}.}
    \label{fig:enter-label}
\end{figure}

Figure~1 presents the unfolded energy spectra for heavy (Si+Fe) and light (H+He+CNO) primary cosmic rays, based on three different hadronic interaction models: QGSJET-II-04 (left), EPOS-LHC (middle), and SIBYLL 2.3d (right).
These spectra have been corrected for shower-to-shower fluctuations using the unfolding procedure described above.
The all-particle spectrum is obtained as the sum of the light and heavy mass components.

For all three models, the energy spectrum of the heavy component exhibits the knee-like feature around 10$^{17}$~eV, characterized by a change in the spectral slope. 
The all-particle spectra shows smoother transition and less pronounced slope changes compared to the heavy components.
In the light-component spectrum, a hardening, i.e., flattening of the spectrum, is observed above about 10$^{17}$~eV, with the spectral slope changing gradually rather than sharply.

The analysis method and procedure introduce systematic uncertainties in the reconstructed energy spectrum and its extracted fit parameters.
Some uncertainties may also be introduced by the reconstruction of the air showers.
All these individual systematic contributions were considered 
and combined in quadrature to obtain the total systematic uncertainty in the flux, which is is of the order of about 13\% for light and 10\% for heavy primaries at the primary energy of $10^{17}$~eV.
Further details on systematic uncertainties are discussed in Ref.~\cite{KANG}.

\begin{figure}[b]
\centering
\includegraphics[width=1.0\linewidth]{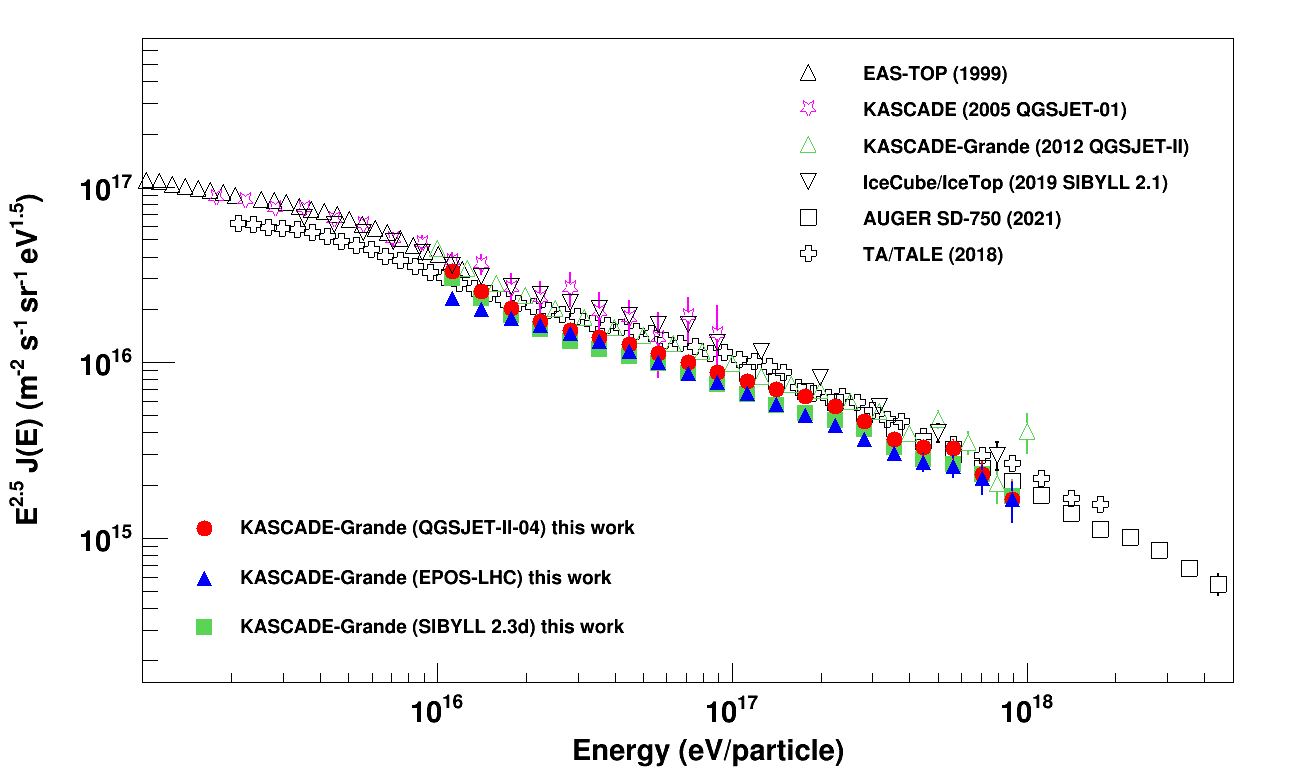}
\caption{
  Comparison of the reconstructed all-particle energy spectrum with results from other experiments: ESA-TOP, IceCube/IceTop, Pierre Auger Observatory, and Telescope Array.
  The markers indicate, for this work, the results using different hadronic interaction models: QGSJET-II-04 (red circle), EPOS-LHC (blue triangles) and SIBYLL 2.3d (green squares), respectively.
}
\end{figure}

Figure~2 compares the resulting all-particle energy spectra derived from different post-LHC interaction models with results from other experiments: ESA-TOP, IceCube/IceTop, Pierre Auger Observatory (PAO) and Telescope Array (TA).
The all-particle energy spectra from KASCADE-Grande are obtained by summing the individual heavy and light component spectra.
As shown in Fig.~2, KASCADE-Grande results on the all-particle spectrum show relatively weak dependence among post-LHC hadronic interaction models in use.
However, the absolute flux measured by KASCADE-Grande is up to 10\% lower than that measured by other experiments. This is possibly due to the observation altitude near sea level, as a different part of the shower evolution is taken into account in the interpretation by the hadronic interaction models.
Furthermore, these differences are related to the absolute normalization of the energy scale by the various models.
Despite differences in observation levels, analysis techniques and hadronic interaction models among the various experiments, there is good agreement in the energy range from PeV to EeV.
At higher energies around $10^{18}$~eV, KASCADE-Grande results are statistically consistent with the result of the Pierre Auger Observatory.

\section{New hadronic interaction model EPOS-LHC-R}

A new version of the EPOS model, EPOS-LHC-R~\cite{EPOSR, EPOSR2025}, has recently been released.
Improvements to the proton-proton cross-sections, multiplicity of proton-proton and proton-Air interactions, and nuclear fragmentation processes are applied.
These changes impact both the shower maximum $X_{max}$ and the muon production in extensive air showers.
An improved hadronization according to LHC data enhances the accuracy of muon production predictions, while the updated cross-section and nuclear fragmentation processes affect $X_{max}$.
Compared to the previous version, EPOS-LHC, the EPOS-LHC-R model predicts an increased number of muons and a deeper $X_{max}$ by approximately 25 g/cm$^2$, and thus even deeper than the predictions from the SIBYLL 2.3d model.
This is expected to lead to an interpretation of a heavier composition for a given measured $X_{max}$.
More details on EPOS-LHC-R are provided in Ref.~\cite{EPOSR, EPOSR2025}.

Full Monte-Carlo simulations, including the detector response of KASCADE-Grande, have been performed, using EPOS-LHC-R.
Figure~3 (left) shows the correlation between the number of charged particles and the number of muons, $y_{CIC}$.
The selection criteria separating heavy and light primaries is based on a fit to the average of silicon and CNO, yielding a separation value of 0.8496, indicated by the dashed line in Fig.~3.
The right panel of Fig.~3 displays the energy calibration functions, described by: ${\rm log}_{10}(E_{true}/{\rm GeV}) = a \cdot {\rm log}_{10}(N_{ch}) + b$, with coefficients $a=0.897$, $b=1.705$ for heavy primaries, and $a=0.953$, $b=1.109$ for light primaries.
The $y_{CIC}$ value for the new EPOS-LHC-R model is slightly lower than that of the previous EPOS model and closer to the values from the QGSJET-II-04 and SIBYLL 2.3d models. This indicates that EPOS-LHC-R predicts a slightly smaller relative muon content in air showers, aligning more closely with the predictions of QGSJET-II-04 and SIBYLL 2.3d models.
The fact that the relative muon content is smaller means that for the observation with KASCADE-Grande the effect of the deeper $X_{max}$ (the electromagnetic component) outweighs the increased generated muon number in the model.

\begin{figure}[t]
    \centering
    \includegraphics[width=0.51\linewidth]{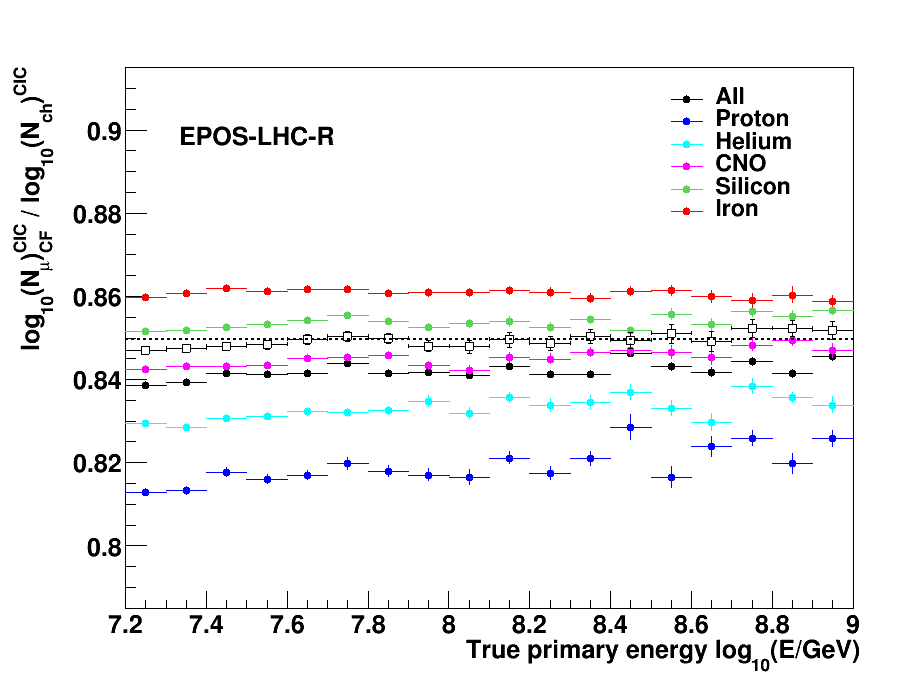}
    \hspace{5 mm}
    \includegraphics[width=0.42\linewidth]{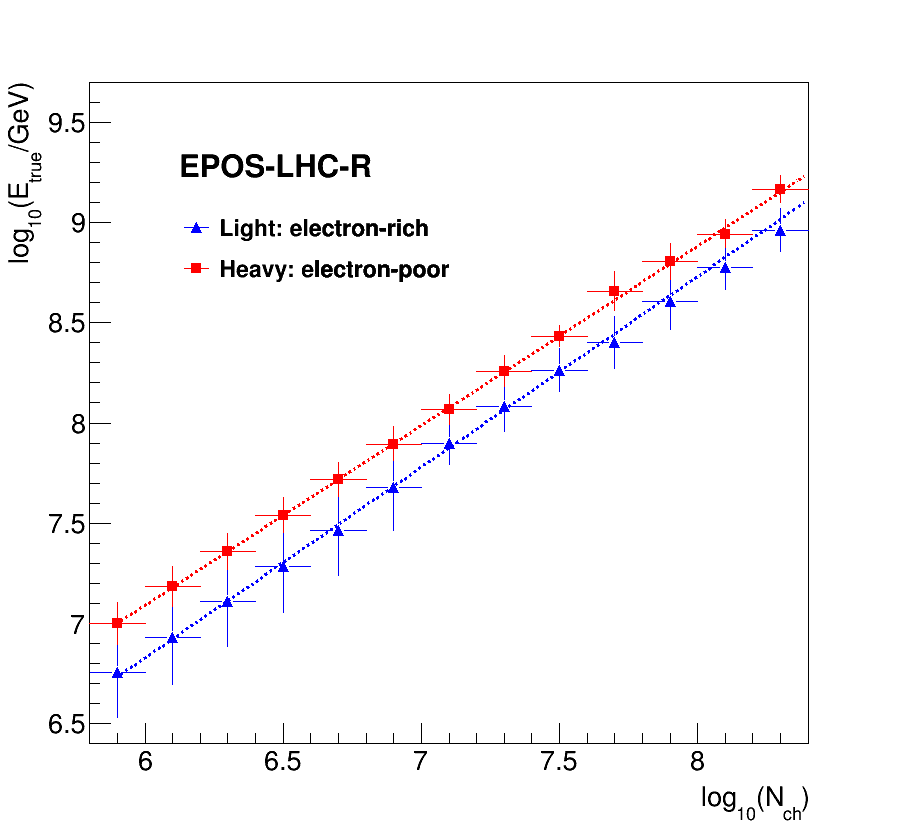}
    \caption{Left: The $y_{CIC}$ parameter as a function of the primary energy for EPOS-LHC-R. The dashed line indicates the selection criteria separating heavy and light mass groups. 
    Right: Primary energy as a function of the number of charged particles for the EPOS-LHC-R model. Linear fits in log-log scale are shown for heavy (red) and light (blue) primary groups.}
    \label{fig:enter-label}
\end{figure}

The reconstructed energy spectra using EPOS-LHC-R are shown in Figure~4 (left), where shower-to-shower fluctuations are not yet taken into account.
A comparison with the EPOS-LHC model (unfolded) is presented in Fig.~4 (right).
For heavy primaries, EPOS-LHC-R shows a slightly higher flux than EPOS-LHC, though spectral features remain similar.
For light primaries, the EPOS-LHC-R model is less dominant compared to EPOS-LHC, resulting in a slightly lower all-particle spectrum for EPOS-LHC-R.
Interestingly, both models show negligible difference above $10^{17}$~eV, while more notable differences appear in the lower energy range.

\begin{figure}[b]
    \centering
    \includegraphics[width=0.49\linewidth]{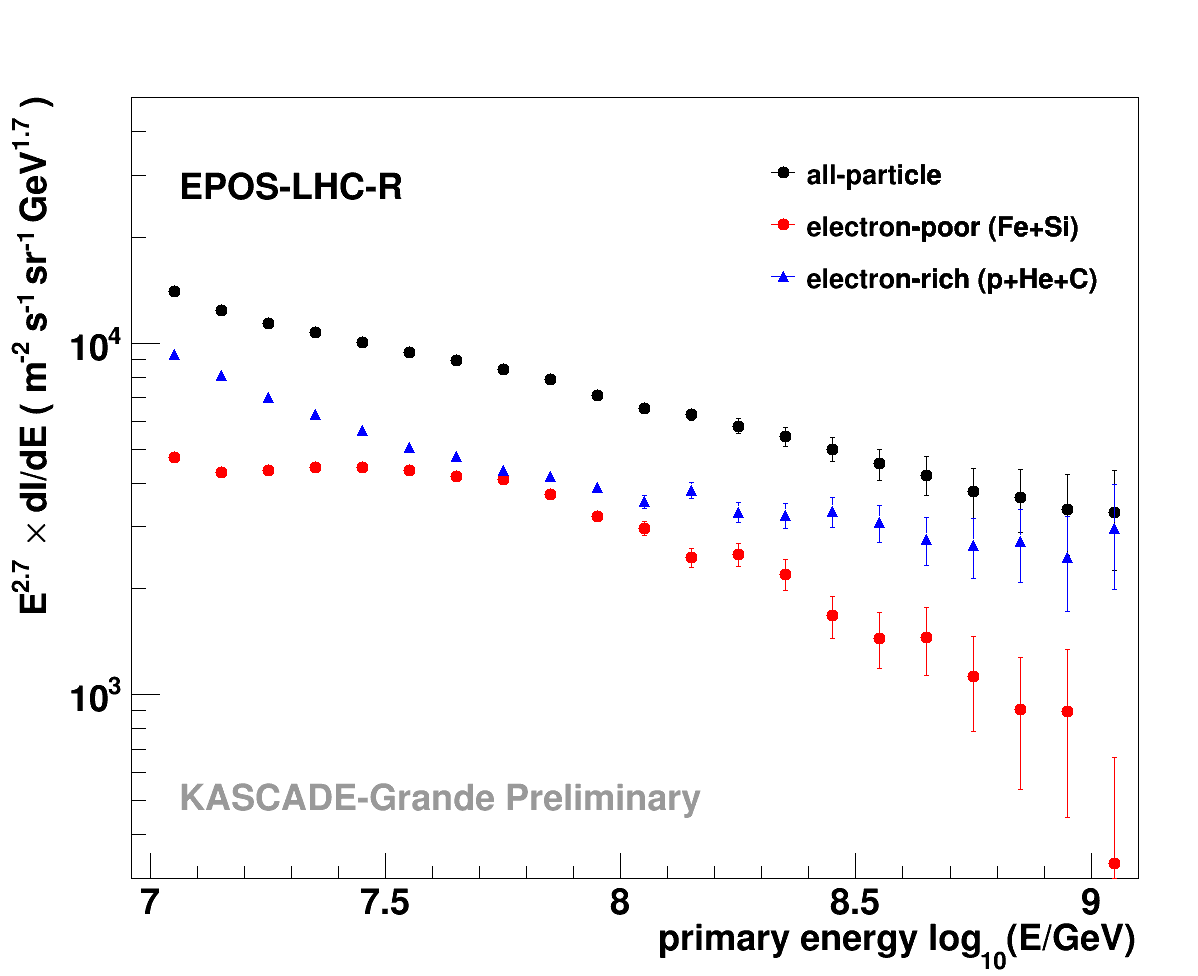}
    \includegraphics[width=0.49\linewidth]{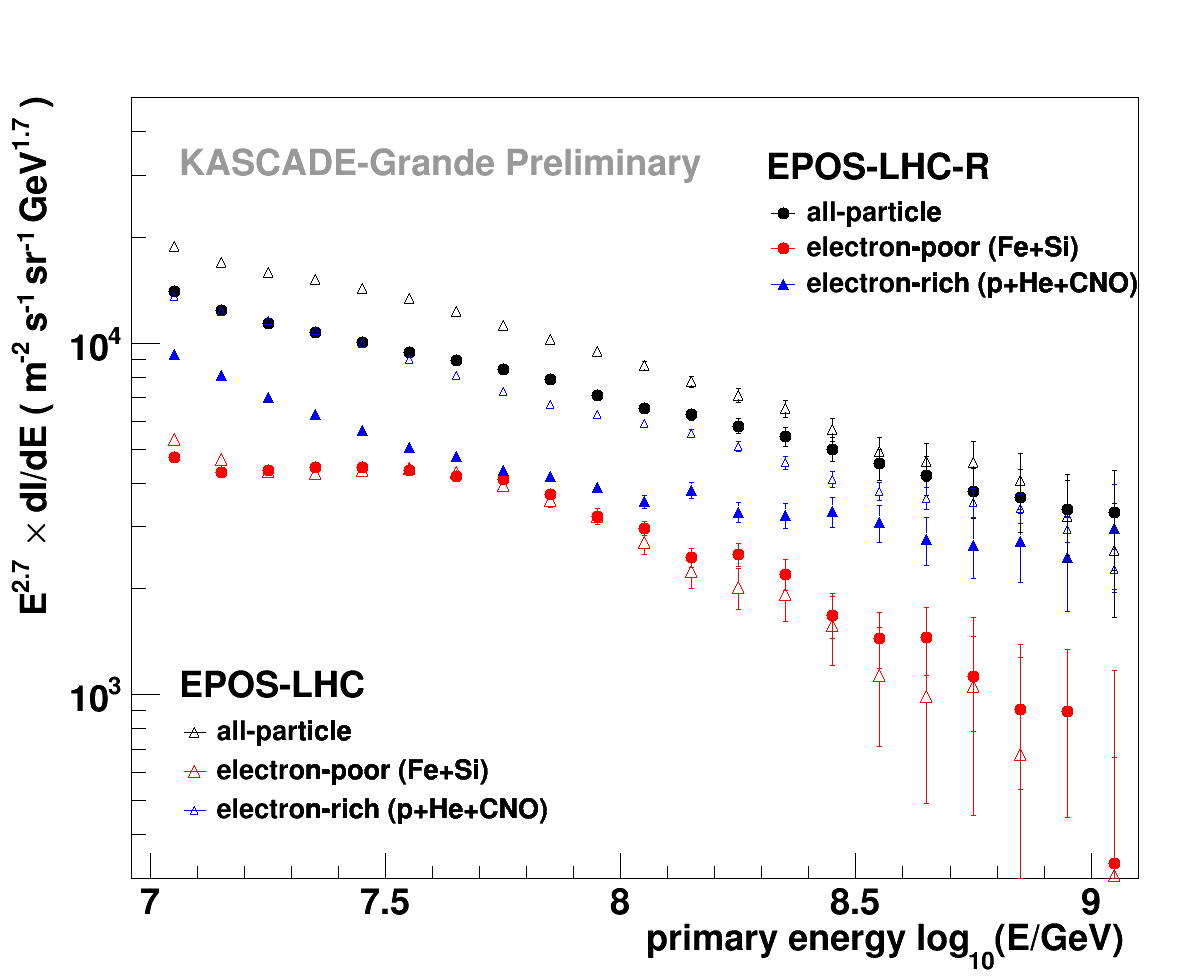}
    \caption{Left: Reconstructed energy spectra for all-particle (black), heavy (red), and light (blue) primaries, based on the EPOS-LHC-R model. The all-particle spectrum is obtained by summing the heavy and light components. Shower fluctuations have not yet been corrected for EPOS-LHC-R.
    Right: Comparison of EPOS-LHC-R (solid markers) with the previous EPOS-LHC model (unfolded) (open markers).}
    \label{fig:enter-label}
\end{figure}

\begin{figure}
    \centering
        \includegraphics[width=0.49\linewidth]{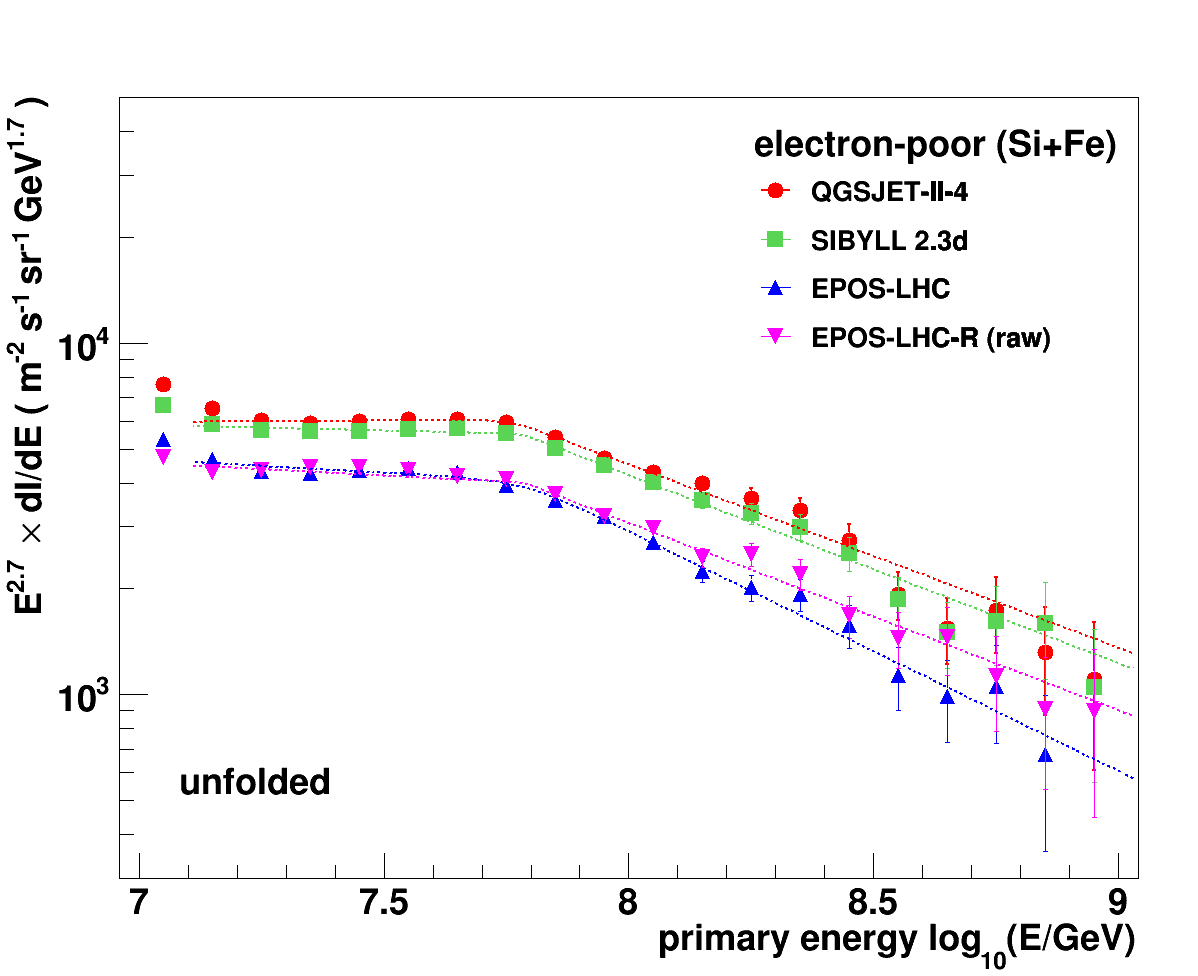}
            \includegraphics[width=0.49\linewidth]{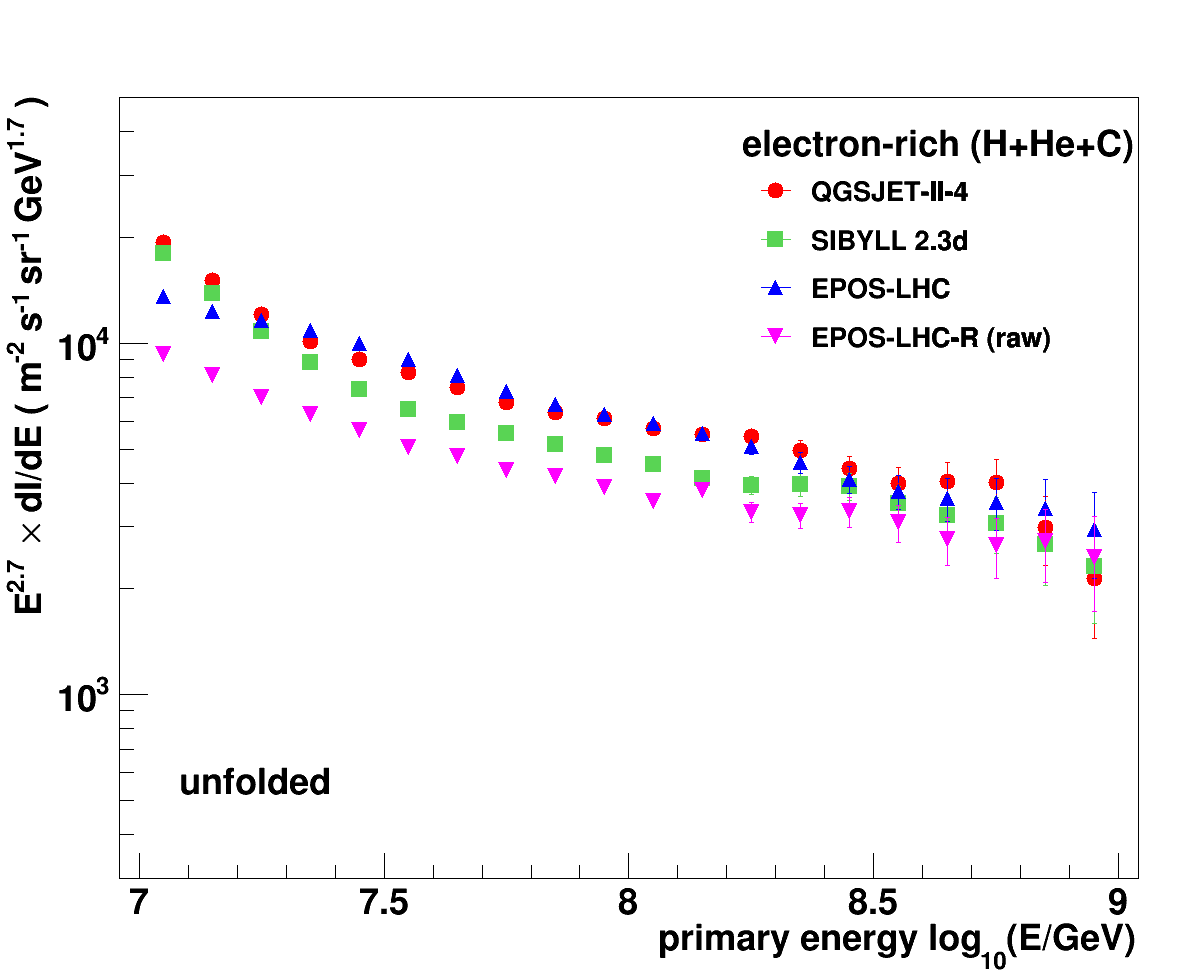}
    \caption{Comparison of the reconstructed energy spectra of heavy (left) and light (right) primaries from EPOS-LHC-R with four different interaction models: QGSJET-II-04, EPOS-LHC, SIBYLL 2.3d, and EPOS-LHC-R. The heavy mass group includes silicon and iron, while the light group contains proton, helium and carbon. Dashed lines represent broken power-law fits.}
    \label{fig:enter-label}
\end{figure}

Figure~5 compares the energy spectra of heavy (Si+Fe) and light (H+He+CNO) mass groups for four different interaction models: QGSJET-II-04, EPOS-LHC, SIBYLL 2.3d, and EPOS-LHC-R.
The first three models show unfolded spectra, whereas the EPOS-LHC-R spectrum is still raw.
The heavy components spectra show similar knee-like structure across all models.
QGSJET-II-04 and SIBYLL 2.3d yield comparable heavy spectra. 
In contrast, both EPOS-LHC and EPOS-LHC-R present slightly lower heavy fluxes, mainly due to differences in the $N_{ch}/N_{\mu}$ ratio.
Light (electron-rich) spectra are consistently more abundant due to the separation boundary near the CNO mass group.
All models exhibit similar hardening of light primaries with a smooth change of the slope around 10$^{17}$~eV.
The EPOS-LHC-R model shows the lowest flux of light primaries of all models, but this probably will slightly increase after applying shower-to-shower fluctuation corrections,
potentially aligning more closely with QGSJET-II-04 and SIBYLL 2.3d.

To quantify the spectral features, a broken power-law fit is applied to the spectra of heavy particles.
The break position and the spectral indices before and after the break are summarized in Table 1.
Spectral breaks appear just below 10$^{17}$~eV, with EPOS-LHC showing the most pronounced slope change.

\begin{table}[h!]
\begin{center}
\small    
\begin{tabular}{lccccc}
\hline
electron-poor & log$_{10}(E_{k}/{\rm GeV})$ & $\gamma_{1}$ & $\gamma_{2}$ & $\Delta \gamma$ & $\chi^{2}$/ndf \\ \hline
QGSJET-II-04  & 7.77 $\pm$ 0.02  & 2.69 $\pm$ 0.01 & 3.22 $\pm$ 0.04 & 0.53 & 3.92 \\
EPOS-LHC      & 7.79 $\pm$ 0.04  & 2.78 $\pm$ 0.01 & 3.38 $\pm$ 0.06 & 0.60 & 4.03 \\
SIBYLL 2.3d   & 7.78 $\pm$ 0.02  & 2.73 $\pm$ 0.01 & 3.24 $\pm$ 0.03 & 0.51 & 1.68 \\
EPOS-LHC-R    & 7.79 $\pm$ 0.01  & 2.77 $\pm$ 0.01 & 3.23 $\pm$ 0.13 & 0.46 & 4.46 \\ \hline
\end{tabular}
\caption{Spectral break positions and spectral indices obtained from broken power-law fits to the unfolded spectra of electron-poor (heavy) primaries. Note that the fit results of the EPOS-LHC-R model are based on the raw (not unfolded) spectrum.}
\end{center}
\end{table}

\section{Conclusion}
Using the shower size measured by KASCADE-Grande and the $y_{CIC}$ technique, the energy spectra of heavy and light mass groups have been reconstructed based on the various post-LHC hadronic interaction models: QGSJET-II-04, EPOS-LHC, SIBYLL 2.3d, and the newly released EPOS-LHC-R.
The EPOS-LHC-R model, tested with KASCADE-Grande data for the first time, show spectral behavior consistent with other models and confirms previous findings. 
To correct bin-to-bin migrations caused by shower-to-shower fluctuations, response matrices to each interaction model were constructed and applied to unfold the reconstructed spectra. 
The impact of this unfolding procedure on the resulting flux is estimated to be less than 10\% for all energy bins and thus does not significantly change the spectral shapes.
All prominent spectral features observed in previous measurements are confirmed: 
observation of a heavy knee at around $10^{17}$~eV accompanied by a flattening of the light component at the same energy. This flattening may be an indication of an extragalactic component, which appears to be dominant already below the energy of $10^{17}$~eV for all post-LHC models.

Additionally, the muon content in cosmic-ray air showers was investigated in the primary energy from $10^{16}$ and $10^{18}$~eV for three zenith angle intervals.
While a slight zenith angle dependence is observed, the measured muon contents by KASCADE-Grande shows a reasonable agreement with the MC expectations. Further details can be found in Ref.~\cite{JuanCarlos}.

Finally, the KASCADE Cosmic-ray Data Centre (KCDC)~\cite{KCDC} provides a public accessible, web-based platform (https://kcdc.iap.kit.edu), where scientific data of the completed KASCADE and KASCADE-Grande (and more) experiments are made available for the astroparticle community and interested public (for latest news on KCDC see also Ref.~\cite{Vika_ICRC25}).

\end{document}